\documentclass[aps,twocolumn,english,superscriptaddress,citeautoscript,
preprintnumbers,amsmath,amssymb,floatfix,footinbib]{revtex4-1}
\usepackage{hyperref}
\hypersetup{colorlinks=true,linkcolor=red,citecolor=blue}
\usepackage{amsmath}
\usepackage{graphicx}
\usepackage{dcolumn}
\usepackage{float}
\usepackage[utf8]{inputenc}

\usepackage{epstopdf}
\usepackage{siunitx}
\usepackage{fancyhdr}
\begin{document}

\title{Damping in yttrium iron garnet film with an interface}

\author{Ravinder Kumar}
\email{ravindk@iitk.ac.in}
\affiliation{Department of Physics, Indian Institute of Technology, Kanpur 208016, India.}
\affiliation{Institute of Physics, Bhubaneswar 751005, India.}
\author{B. Samantaray}
\affiliation{Department of Physics, Indian Institute of Technology, Kanpur 208016, India.}
\author{Shubhankar Das}
\affiliation{Institute of Physics, Johannes Gutenberg-University Mainz, 55099 Mainz, Germany.}
\author{Kishori Lal}
\affiliation{Department of Physics, Indian Institute of Technology, Kanpur 208016, India.}
\author{D. Samal}
\affiliation{Institute of Physics, Bhubaneswar 751005, India.}
\affiliation{Homi Bhabha National Institute, Anushakti Nagar, Mumbai 400085, India.}
\author{Z. Hossain}
\email{zakir@iitk.ac.in}
\affiliation{Department of Physics, Indian Institute of Technology, Kanpur 208016, India.}
\affiliation{Institute of Low Temperature and Structure Research, 50-422 Wroclaw, Poland.}

\date{\today}
             
\begin{abstract}

 We report strong damping enhancement in a 200 nm thick yttrium iron garnet (YIG) film due to spin inhomogeneity at the interface. The growth-induced thin interfacial gadolinium iron garnet (GdIG) layer antiferromagnetically (AFM) exchange couples with the rest of the YIG layer. The out-of-plane angular variation of ferromagnetic resonance (FMR) linewidth $\Delta H$ reflects a large inhomogeneous distribution of effective magnetization $\Delta 4 \pi M_{eff}$ due to the presence of an exchange springlike moments arrangement in YIG. We probe the spin inhomogeneity at the YIG-GdIG interface by performing an in-plane angular variation of resonance field $H_{r}$, leading to a unidirectional feature. The large extrinsic $\Delta 4\pi M_{eff}$ contribution, apart from the inherent intrinsic Gilbert contribution, manifests enhanced precessional damping in YIG film. 

\vskip 1cm
\end{abstract}

\maketitle
\newpage

\section{Introduction}

The viability of spintronics demands novel magnetic materials and YIG is a potential candidate as it exhibits ultra-low precessional damping, $\alpha\sim 3\times10^{-5}$\cite{Sparks}. The magnetic properties of YIG thin films epitaxially grown on top of Gd$_{3}$Ga$_{5}$O$_{12}$ (GGG) vary significantly due to growth tuning\cite{YIG_Damping, Kumar}, film thickness\cite{YIG_YAG}, heavy metals substitution\cite{BiYIG_MO, Kumar_2019, Onbasli2016} and coupling with thin metallic layers\cite{YIG_Pt_Damping_2013, YIG_AFM_NM_2015, Chang2017}. The growth processes may also induce the formation of a thin interfacial-GdIG layer at the YIG-GGG interface\cite{APLEva, Synthetic-antiferro_PRAppl, EB_IHL_YIG_2021}. The YIG-GdIG heterostructure derived out of monolithic YIG film growth on GGG exhibits interesting phenomena such as all-insulating equivalent of a synthetic antiferromagnet\cite{Synthetic-antiferro_PRAppl} and hysteresis loop inversion governed by positive exchange-bias \cite{EB_IHL_YIG_2021}. The radio frequency magnetization dynamics on YIG-GdIG heterostructure still remains unexplored and need a detailed FMR study.

The relaxation of magnetic excitation towards equilibrium is governed by intrinsic and extrinsic mechanisms, leading to a finite $\Delta H$\cite{LW_Damping_2003, LW_Damping_2007}. The former mechanism dictates Gilbert type relaxation, a consequence of direct energy transfer to the lattice governed by both spin-orbit coupling and exchange interaction in all magnetic materials\cite{LW_Damping_2003, LW_Damping_2007}. Whereas, the latter mechanism is a non-Gilbert-type relaxation, divided mainly into two categories\cite{LW_Damping_2003, LW_Damping_2007}- (i) the magnetic inhomogeneity induced broadening: inhomogeneity in the internal static magnetic field, and the crystallographic axis orientation; (ii) two-magnon scattering: the energy dissipates in the spin subsystem by virtue of magnon scattering with nonzero wave vector, $k \neq 0$, where, the uniform resonance mode couples with the degenerate spin waves. The angular variation of $H_{r}$ provides information about the presence of different magnetic anisotropies\cite{YIG_YAG, Kumar_2019}. Most attention has been paid towards the angular dependence of $H_{r}$\cite{YIG_YAG, Kumar_2019}, whereas, the angular variation of the $\Delta H$ is sparsely investigated. The studies involving angular dependence of $\Delta H$ may help to probe different contributions to the precessional damping. 

In this paper, the effects of intrinsic and extrinsic relaxation mechanisms on precessional damping of YIG film is studied extensively using FMR technique. An enhanced value of $\alpha \sim 1.2 \times 10^{-3}$ is realized, which is almost two orders of magnitude higher than what is usually seen in YIG thin films, $\sim 6 \times 10^{-5}$\cite{Sparks, YIG_Damping}. The out-of-plane angular variation of $\Delta H$ shows an unusual behaviour where spin inhomogeneity at the interface plays significant role in defining the $\Delta H$ broadening and enhanced $\alpha$. In-plane angular variation showing a unidirectional feature, demands the incorporation of an exchange anisotropy to the free energy density, evidence of the presence of an AFM exchange coupling at the YIG-GdIG interface. The AFM exchange coupling leads to a Bloch domain-wall-like spiral moments arrangement in YIG and gives rise to a large $\Delta 4 \pi M_{eff}$. This extrinsic $\Delta 4 \pi M_{eff}$ contribution due to spin inhomogeneity at the interface adds up to the inherent Gilbert contribution, which may lead to a significant enhancement in precessional damping.

\section{Sample and Measurement Setups}
\label{sec:Sample and Measurement Setups}

We deposit a $\sim 200$ nm thick epitaxial YIG film on GGG(111)-substrate by employing a KrF Excimer laser (Lambda Physik COMPex Pro, $\lambda$ = 248 \si{\nano\meter}) of 20 \si{\nano\second} pulse width. A solid state synthesized $Y_{3}Fe_{5}O_{12}$ target is ablated using an areal energy of 2.12 J.cm$^{-2}$ with a repetition frequency of 10 \si{\Hz}. The GGG(111) substrate is placed 50 \si{\milli\meter} away from the target. The film is grown at 800 $^{o}$C temperature and {\it in-situ} post annealed at the same temperature for 60 minutes in pure oxygen environment. The $\theta-2\theta$ X-ray diffraction pattern shows epitaxial growth with trails of Laue oscillations (Fig. 3(a) of ref\cite{Kumar}). FMR measurements are performed using a Bruker EMX EPR spectrometer and a broadband coplanar waveguide (CPW) setup. The former technique uses a cavity mode frequency $f \approx 9.60$ GHz, and enables us to perform FMR spectra for various $\theta_{H}$ and $\phi_{H}$ angular variations. The latter technique enables us to measure frequency dependent FMR spectra. We define the configurations $\textbf{H}$ parallel ($\theta_{H} = 90^{o}$) and perpendicular ($\theta_{H} = 0^{o}$) to the film plane for rf frequency and angular dependent measurements. The resultant spectra are obtained as the derivative of microwave absorption w.r.t. the applied field $\textbf{H}$.

\section{Results and Discussion}  
\label{sec:Results and Discussion}

\subsection{Broadband FMR}
\label{subsec:Broadband FMR}

\begin{figure*}[t]
\begin{center}
\includegraphics[trim =0mm 0mm 0mm 0mm, width=170mm]{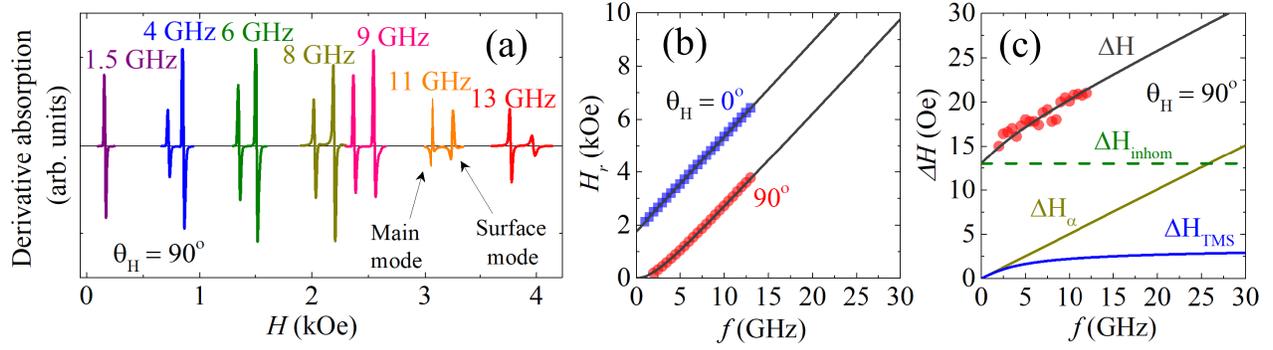}
\setlength{\belowcaptionskip}{-8mm}
\caption{Room temperature frequency dependent FMR measurements. (a) Representative FMR derivative spectra for different frequencies at $\theta_{H} = 90^{o}$. (b) Resonance field vs. frequency data for $\theta_{H} = 90^{o}$ and $\theta_{H}= 0^{o}$ are represented using red and blue data points, respectively. The fitting to both the data are shown using black lines. (c) Linewidth vs. frequency data at $\theta_{H} = 90^{o}$. The solid red circles represent experimental data, whereas the solid black line represents $\Delta H$ fitting. Inhomogeneous ($\Delta H_{inhom}$), Gilbert ($\Delta H_{\alpha}$) and two-magnon scattering ($\Delta H_{TMS}$) contributions to $\Delta H$ are shown using dashed green, solid yellow and blue lines, respectively.}\label{Fig:Broadband FMR}
\end{center}
\end{figure*}

Fig. \ref{Fig:Broadband FMR}(a) shows typical broadband FMR spectra in a frequency $f$ range of 1.5 to 13 \si{\giga\Hz} for 200 nm thick YIG film at temperature $T=300$ K and $\theta_{H}=90^{o}$. The mode appearing at a lower field value is the main mode, whereas the one at higher field value represents surface mode. We discuss all these features in detail in the succeeding subsection \ref{subsec:Cavity FMR}. We determine the resonance field $H_{r}$ and linewidth (peak-to-peak linewidth) $\Delta H$ from the first derivative of the absorption spectra. Fig. \ref{Fig:Broadband FMR}(b) shows the rf frequency dependence of $H_{r}$ at $\theta_{H} = 90^{o}$ and $0^{o}$. We use the Kittel equation for fitting the frequency vs. $H_{r}$ data from the resonance condition expressed as\cite{Chang2017}, $f = \gamma \left[ {H_{r}} ({H_{r}} + 4\pi {M_{eff}}) \right]^{1/2}/(2{\pi})$ for $\theta_{H}=90^{o}$ and $f = \gamma ({H_{r}} - 4\pi {M_{eff}})/(2{\pi})$ for $\theta_{H}=0^{o}$. Where, $\gamma=g\mu_{B}/\hslash$ is the gyromagnetic ratio, $4\pi M_{eff}= 4\pi M_{S}-H_{ani} $ is the effective magnetization consisting of $4\pi M_{S}$ saturation magnetization (calculated using M(H)) and $H_{ani}$ anisotropy field parametrizing cubic and out-of-plane uniaxial anisotropies. The fitting gives $4\pi M_{eff} \approx 2000$ Oe, which is used to calculate the $H_{ani} \approx -370$ Oe.

  Fig. \ref{Fig:Broadband FMR}(c) shows the frequency dependence of $\Delta H$ at $\theta_{H}=90^{o}$. The intrinsic and extrinsic damping contributions are responsible for a finite width of the FMR signal. The intrinsic damping $\Delta H_{int}$ arises due to the Gilbert damping of the precessing moments. Whereas, the extrinsic damping $\Delta H_{ext}$ exists due to different non-Gilbert-type relaxations such as inhomogeneity due to the distribution of magnetic anisotropy $\Delta H_{inhom}$, or two-magnon scattering (TMS) $\Delta H_{TMS}$. The intrinsic Gilbert damping coefficient ($\alpha$) can be determined using the Landau-Liftshitz-Gilbert equation expressed as\cite{Chang2017}, $\Delta H = \Delta H_{in} + \Delta H_{inhom} = (4\pi\alpha/\sqrt{3}\gamma)f + \Delta H_{inhom}$. Considering the above equation where $\Delta H$ obeys linear $f$ dependence, the slope determines the value of $\alpha$, and $\Delta H_{inhom}$ corresponds to the intercept on the vertical axis. We observe a very weak non-linearity in the $f$ dependence of $\Delta H$, which is believed to be due to the contribution of TMS to the linewidth $\Delta H_{TMS}$. The non-linear $f$ dependence of $\Delta H$ in Fig. \ref{Fig:Broadband FMR}(c) can be described in terms of TMS, assuming $\Delta H = \Delta H_{in} + \Delta H_{inhom} + \Delta H_{TMS}$. We put a factor of $1/\sqrt{3}$ to $\Delta H$ due to the peak-to-peak linewidth value extraction\cite{LW_Damping_2003}. The TMS induces non-linear slope at low frequencies, whereas a saturation is expected at high frequencies. TMS is induced by scattering centers and surface defects in the sample. The defects with size comparable to the wavelength of spin waves are supposed to act as scattering centres. The TMS term at $\theta_{H} = 90^{o}$ can be expressed as\cite{Impurity_relaxation_2017}-

\begin{equation}
\label{Eq: IP TMS}
 \Delta H_{TMS}(\omega) = \Gamma \sin^{-1}  \sqrt{\frac{\sqrt{\omega^{2}+(\omega_{0}/2)^{2}}-\omega_{0}/2}{\sqrt{\omega^{2}+(\omega_{0}/2)^{2}}+\omega_{0}/2}},
\end{equation}

with $\omega = 2 \pi f$ and $\omega_{0} = \gamma 4 \pi M_{eff}$. The prefactor $\Gamma$ defines the strength of TMS. The extracted values are as follows: $\alpha = 1.2 \times 10^{-3}$, $\Delta H_{0} = 13$ Oe and $\Gamma = 2.5$ Oe. The Gilbert damping for even very thin YIG film is extremely low, $\sim 6 \times 10^{-5}$. Whereas, the value we achieved is higher than the reported in the literature for YIG thin films\cite{YIG_Damping}. Also, the value of $\Gamma$ is insignificant, implying negligible contribution to the damping.

\subsection{Cavity FMR}
\label{subsec:Cavity FMR}

\begin{figure*}[t]
\begin{center}
\includegraphics[trim = 0mm 0mm 0mm 0mm, width=170mm]{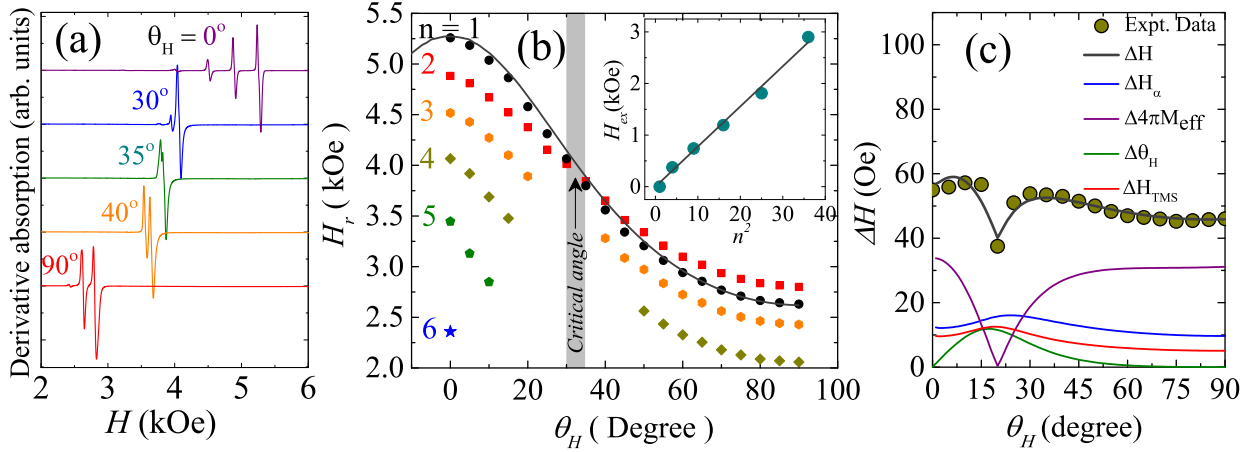}
\setlength{\belowcaptionskip}{-8mm}
\caption{Room temperature out-of-plane angular $\theta_{H}$ dependence of FMR. (a) Derivative FMR spectra shown for different $\theta_{H}$ performed at $\approx 9.6$ GHz. (b) $\theta_{H}$ variation of uniform mode and SWR modes of resonance field $H_{r}$. Inset: Exchange field ($H_{ex}$) vs mode number square ($n^{2}$). (c) $\theta_{H}$ variation of the linewidth ($\Delta H$), where, the experimental and simulated data are represented by solid yellow circles and black line, respectively. The different contributions $\Delta H_{\alpha}$, $\Delta 4 \pi M_{eff}$, $\Delta \theta_{H}$ and $\Delta H_{TMS}$ are represented by gray, purple, green and red lines, respectively.}\label{Fig:Angular_FMR}
\end{center}
\end{figure*}

Fig. \ref{Fig:Angular_FMR}(a) shows typical $T = 300$ K cavity-FMR ($f \approx 9.6$ GHz) spectra for YIG film performed at different $\theta_{H}$. The FMR spectra exhibit some universal features: (i) Spin-Wave resonance (SWR) spectrum for $\theta_{H} = 0^{o}$; (ii) rotating the $\textbf{H}$ away from the $\theta_{H} = 0^{o}$, the SWR modes successively start diminishing, and at certain critical angle $\theta_{c}$ (falls in a range of $30-35^{o}$; shaded region in Fig. \ref{Fig:Angular_FMR}(b)), all the modes vanish except a single mode (uniform FMR mode). Further rotation of $\textbf{H}$ for $\theta_{H}>\theta_{c}$, the SWR modes start re-emerging. We observe that the SWR mode appearing at the higher field side for $\theta_{H}>\theta_{c}$, represents an exchange-dominated non-propagating surface mode\cite{Surface_Anisotropy_1977, SWR_Critical_Angle_YIG, SWR_Critical_Angle_2007}. The above discussed complexity in $H_{r}$ vs $\theta_{H}$ behaviour has already been realized in some material systems\cite{SWR_Critical_Angle_2007}, including a $\mu$-thick YIG film\cite{SWR_Critical_Angle_YIG}. The localized mode or surface spin-wave mode appears for $\textbf{H} \parallel$ but not $\perp$ to the film-plane\cite{Surface_Anisotropy_1977, SWR_Critical_Angle_YIG, SWR_Critical_Angle_2007}. We assign the SWR modes for the sequence $n = 1,2,3,....$, as it provides the best correspondence to $H_{ex} \propto n^{2}$, where, $H_{ex} = H_{r}(n) - H_{r}(0)$ defines exchange field\cite{Exchange_Stiffness_YIG_2014}. The exchange stiffness can be obtained by considering the modified Schreiber and Frait classical approach using the mode number $n^{2}$ dependence of resonance field (inset Fig. 3(c))\cite{Exchange_Stiffness_YIG_2014}. For a fixed frequency, the exchange field $H_{ex}$ of thickness modes is determined by subtracting the highest field resonance mode ($n = 1$) from the higher modes ($n \neq 1$). In modified Schreiber and Frait equation, the $H_{ex}$ shows direct dependency on the exchange stiffness $D$: $\mu_{0} H_{ex} = D \frac{\pi^{2}}{d^{2}} n^{2}$ (where $d$ is the film thickness). The linear fit of data shown in the inset of Fig. \ref{Fig:Angular_FMR}(b) gives $D = 3.15 \times 10^{-17}$ T.m$^{2}$. The exchange stiffness constant $A$ can be determined using the relation $A = D\ M_{S}/2$. The calculated value is $A = 2.05$ pJ.m$^{-1}$, which is comparable to the value calculated for YIG, $A = 3.7$ pJ.m$^{-1}$\cite{Exchange_Stiffness_YIG_2014}.

YIG thin films with in-plane easy magnetization exhibit extrinsic uniaxial magnetic and intrinsic magnetocrystalline cubic anisotropies\cite{Lee2016_JAP}. The total free energy density for YIG(111) is given by\cite{Landau, Lee2016_JAP}:

\begin{equation}
\label{Eq:Free energy density}
\begin{array}{c}
F =  - H{M_S}\left[ \begin{array}{l}\sin {\theta _H}\sin {\theta _M}\cos\left( {{\phi _H} - {\phi _M}} \right)\\ + \cos {\theta _H}\cos {\theta _M}\end{array} \right] \\
+ 2\pi M_S^2{\cos ^2}{\theta _M} - {K_u}{\cos ^2}{\theta _M} \\
+ \frac{{{K_1}}}{{12}}\left( \begin{array}{l} 7{\sin ^4}{\theta _M} - 8{\sin ^2}{\theta _M} + 4 - \\ 4\sqrt 2 {\sin ^3}{\theta _M}\cos {\theta _M}\cos 3{\phi _M}
\end{array} \right)\\
 + \frac{{K_2}}{{108}}\left(\begin{array}{l} 
 - 24{{\sin }^6}{\theta _M} + 45{{\sin }^4}{\theta _M} - 24{{\sin }^2}{\theta _M} + 4 \\
- 2\sqrt 2 {{\sin }^3}{\theta _M}\cos {\theta _M}\left( {5{{\sin }^2}{\theta _M} - 2} \right)\cos 3{\phi _M} \\
+ {{\sin }^6}{\theta _M}\cos 6{\phi _M} \end{array} \right)
\end{array}\
\end{equation}

The Eq. \ref{Eq:Free energy density} consists of the following different energy terms; the first term is the Zeeman energy, the second term is the demagnetization energy, the third term is the out-of-plane uniaxial magnetocrystalline anisotropy energy $K_{u}$, and the last two terms are the first and second order cubic magnetocrystalline anisotropy energies ($K_{1}$ and $K_{2}$), respectively. The total free energy density equation is minimized by taking partial derivatives w.r.t. to $\theta_{M}$ and $\phi_{M}$ to obtain the equilibrium orientation of the magnetization vector $\bf{M(H)}$, i.e., $\partial F/\partial \theta_{M} = \partial F/\partial \phi_{M} = 0$. The resonance frequency of uniform precession at equilibrium condition is expressed as\cite{Lee2016_JAP, Suhl, Smit}:

\begin{equation}\label{Eq:Resonance condition}
{\omega _{res}} = \frac{\gamma }{{{M_S}\sin {\theta _M}}}{\left[ {\frac{{{\partial ^2}F}}{{\partial \theta _M^2}}\frac{{{\partial ^2}F}}{{\partial \phi _M^2}} - {{\left( {\frac{{{\partial ^2}F}}{{\partial {\theta _M}\partial {\phi _M}}}} \right)}^2}} \right]^{{\raise0.7ex\hbox{$1$} \!\mathord{\left/
 {\vphantom {1 2}}\right.\kern-\nulldelimiterspace}
\!\lower0.7ex\hbox{$2$}}}}\
\end{equation}

 Mathematica is used to numerically solve the resonance condition described by Eq. \ref{Eq:Resonance condition} for the energy density given by Eq. \ref{Eq:Free energy density}. The solution for a fixed frequency is used to fit the angle dependent resonance data ($H_{r}$ vs. $\theta_{H}$) shown in Fig. \ref{Fig:Angular_FMR}(b). The main mode data simulation is shown using a black line. The parameters obtained from the simulation are $K_{u}= -1.45 \times 10^{4}$ erg.cm$^{-3}$, $K_{1}= 1.50 \times 10^{3}$ erg.cm$^{-3}$, and $K_{2} = 0.13 \times 10^{3}$ erg.cm$^{-3}$. The calculated uniaxial anisotropy field value is $H_{u} \sim -223$ Oe.

 The $\Delta H$ manifests the spin dynamics and related relaxation mechanisms in a magnetic system.  The intrinsic contribution to $\Delta H$ arises due to Gilbert term $\Delta H_{int} \approx \Delta H_{\alpha}$, whereas, the extrinsic contribution $\Delta H_{ext}$ consists of line broadening due to $\Delta H_{inhom}$ and $\Delta H_{TMS}$. The terms representing the precessional damping due to intrinsic and extrinsic contributions can be expressed in different phenomenological forms. Figure \ref{Fig:Angular_FMR}(c) shows $\Delta H$ as a function of $\theta_{H}$. The $\theta_{H}$ variation of $\Delta H$ shows distinct signatures due to different origins of magnetic damping. We consider both $\Delta H_{int}$ and $\Delta H_{ext}$ magnetic damping contributions to the broadening of $\Delta H$, $\Delta H = \Delta H_{\alpha} + \Delta H_{inhom} + \Delta H_{TMS}$. The first term can be expressed as\cite{LW_Damping_2003}-

\begin{equation}
\label{Eq:Gilbert term}
\Delta H_{\alpha} = \frac{\alpha}{M_{S}} \left[ \frac{\partial^{2}F}{\partial\theta^{2}_{M}} + \frac{1}{\sin^{2}\theta_{M}} \frac{\partial^{2}F}{\partial \phi^{2}_{M}} \right] \bigg| \frac{\partial(\frac{2\pi f}{\gamma})}{\partial H_{r}} \bigg|^{-1}.
\end{equation}

The second term $\Delta H_{inhom}$ has a form\cite{LW_Damping_2003}-

\begin{equation}
\label{Eq:Extrinsic}
\Delta H_{inhom} = \bigg| \frac{dH_{r}}{d4\pi M_{eff}} \bigg| \Delta 4\pi M_{eff} + \bigg| \frac{dH_{r}}{d\theta_{H}} \bigg| \Delta\theta_{H}.
\end{equation}

Where, the dispersion of magnitude and direction of the $4\pi M_{eff}$ are represented by $\Delta 4 \pi M_{eff}$ and $\Delta \theta_{H}$, respectively. The $\Delta H_{inhom}$ contribution arises due to a small spread of the sample parameters such as thickness, internal fields, or orientation of crystallites within the thin film. The third term $\Delta H_{TMS}$ can be written as\cite{TMS_2013}-

\begin{equation}
\label{Eq: OOP TMS}
\begin{array}{c}
\Delta H_{TMS} = \sum\limits_{i=1} \frac{\Gamma_{i}^{out} f_{i}(\phi_{H})}{\mu_{0}\gamma\Phi} \sin^{-1} \sqrt{\frac{\sqrt{\omega^{2}+(\omega_{0}/2)^{2}}-\omega_{0}/2}{\sqrt{\omega^{2}+(\omega_{0}/2)^{2}}+\omega_{0}/2}} , \\
\Gamma_{i}^{out} = \Gamma_{i}^{0} \Phi A (\theta - \pi/4) \frac{dH_{r}(\theta_{H})}{d\omega(\theta_{H})} \bigg/ \frac{dH_{r}(\theta_{H}=0)}{d\omega(\theta_{H}=0)}
\end{array}
\end{equation}

The prefactor $\Gamma^{out}_{i}$ defines the TMS strength and has a $\theta_{H}$ dependency in this case. The type and size of the defects responsible for TMS is difficult to characterize which makes it non-trivial to express the exact form of $\Gamma^{out}_{i}$. Although, it may have a simplified expression given in Eq. \ref{Eq: OOP TMS}, where, $\Gamma^{0}_{i}$ is a constant; $A(\theta-\pi/4)$, a step function which makes sure that the TMS is deactivated for $\theta_{H}<\pi/4$; and $dH_{r}(\theta_{H})/d\omega(\theta_{H})$, a normalization factor responsible for the $\theta_{H}$ dependence of the $\Gamma^{out}_{i}$. In fig. \ref{Fig:Angular_FMR}(c)the solid dark yellow circles and black solid line represent the experimental and simulated $\Delta H$ vs $\theta_{H}$ data, respectively. We also plot contributions of different terms such as $\Delta H_{\alpha}$ (blue color line), $\Delta 4 \pi M_{eff}$ (purple color line),  $\Delta \theta_{H}$ (green color line) and $\Delta H_{TMS}$ (red color line). The fitting provides following extracted parameters, $\alpha = 1.3 \times 10^{-3}$, $\Delta 4 \pi M_{eff} = 58$ Oe, $\Delta \theta_{H} = 0.29^{o}$ and $\Gamma_{i}^{0} = 1.3$ Oe. The precessional damping calculated from the $\Delta H$ vs. $\theta_{H}$ corroborate with the value extracted from the frequency dependence of $\Delta H$ data (shown in Fig. \ref{Fig:Broadband FMR}(c)); $\alpha = 1.2 \times 10^{-3}$. The $\Delta H$ broadening and the overwhelmingly enhanced precessional damping are the direct consequence of contributions from intrinsic and extrinsic damping. Usually, the Gilbert term and the inhomogeneity due to sample quality contribute to the broadening of $\Delta H$ and enhanced $\alpha$ in YIG thin films. If we interpret the $\Delta H$ vs $\theta_{H}$ data, it is clear that damping enhancement in YIG is arising from the extrinsic magnetic inhomogeneity.

The role of an interface in YIG coupled with metals or insulators leading to the increments in $\Delta H$ and $\alpha$ has been vastly explored. Wang et. al. \cite{YIG_AFM_NM_2015} studied a variety of insulating spacers between YIG and Pt to probe the effect on spin pumping efficiency. Their results suggest the generation of magnetic excitations in the adjacent insulating layers due to the precessing magnetization in YIG at resonance. This happens either due to fluctuating correlated moments or antiferromagnetic ordering, via interfacial exchange coupling, leading to $\Delta H$ broadening and enhanced precessional damping of the YIG\cite{YIG_AFM_NM_2015}. The impurity relaxation mechanism is also responsible for $\Delta H$ broadening and enhanced magnetic damping in YIG, but is prominent only at low temperatures\cite{Impurity_relaxation_2017}. Strong enhancement in magnetic damping of YIG capped with Pt has been observed by Sun et. al. \cite{YIG_Pt_Damping_2013}. They suggest ferromagnetic ordering in an atomically thin Pt layer due to proximity with YIG at the YIG-Pt interface, dynamically exchange couples to the spins in YIG\cite{YIG_Pt_Damping_2013}. In recent years, some research groups have reported the presence of a thin interfacial layer at the YIG-GGG interface\cite{APLEva,Synthetic-antiferro_PRAppl,EB_IHL_YIG_2021}. The 200 nm film we used in this study is of high quality with a trails of sharp Laue oscillations [see Fig 3(a) in ref.\cite{Kumar}]. Thus it is quite clear that the observed $\Delta H$ broadening and enhanced $\alpha$ is not a consequence of sample inhomogeneity. The formation of an interfacial GdIG layer at the YIG-GGG interface, which exchange couples with the YIG film may lead to $\Delta H$ broadening and increased $\alpha$. Considering the above experimental evidences leading to $\Delta H$ broadening and enhanced Gilbert damping due to coupling with metals and insulators\cite{YIG_AFM_NM_2015, YIG_Pt_Damping_2013}, it is safe to assume that the interfacial GdIG layer at the interface AFM exchange couples with the YIG\cite{APLEva,Synthetic-antiferro_PRAppl,EB_IHL_YIG_2021}, and responsible for enhanced $\Delta H$ and $\alpha$.

\begin{figure}[t]
\begin{center}
\includegraphics[trim = 0mm 0mm 0mm 0mm, width=65mm]{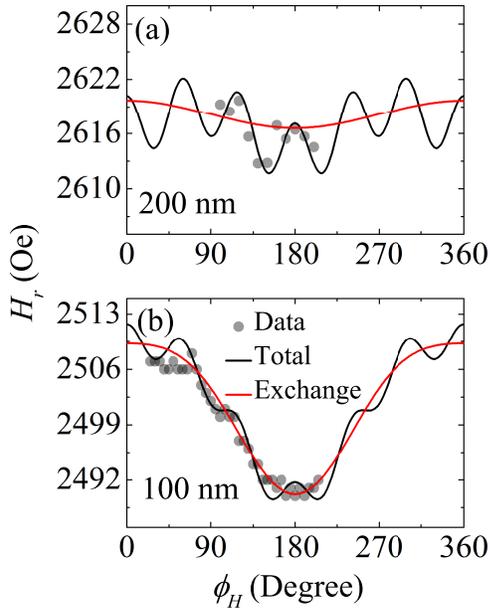}
\setlength{\belowcaptionskip}{-8mm}
\caption{(a) In-plane angular $\phi_{H}$ variation of $H_{r}$. The experimental data are represented by solid grey circles. Whereas, the simulated data for total and exchange (unidirectional) anisotropy are represented by black and red solid lines, respectively. (a) 200 nm thick YIG sample. (b) 100 nm thick YIG sample.}\label{Fig:IP_Hr_Phi_H}
\end{center}
\end{figure}

Fig. \ref{Fig:IP_Hr_Phi_H} shows in-plane $\phi_{H}$ angular variation of $H_{r}$. We simulate the in-plane $H_{r}$ vs $\phi_{H}$ angular variation using the free energy densities provided in ref. \cite{Wang2017} and an additional term, $-{K_{EA}}.{\sin}{\theta _M}.{\cos}{\phi _M}$, representing the exchange anisotropy ($K_{EA}$). Even though $\phi_{H}$ variation of $H_{r}$ shown in Fig. \ref{Fig:IP_Hr_Phi_H}(a) is not so appreciable as the film is 200 nm thick, a very weak unidirectional anisotropy trend is visible, suggesting an AFM exchange coupling between the interface and YIG. It has been shown that the large inhomogeneous $4 \pi M_{eff}$ is a direct consequence of the AFM exchange coupling at the interface of LSMO and a growth induced interfacial layer\cite{IHL_LSMO}. The YIG thin film system due to the presence of a hard ferrimagnetic GdIG interfacial layer possesses AFM exchange coupling\cite{APLEva,Synthetic-antiferro_PRAppl,EB_IHL_YIG_2021}. A Bloch domain-wall-like spiral moments arrangement takes place due to the AFM exchange coupling across the interfacial GdIG and top bulk YIG layer\cite{APLEva,Synthetic-antiferro_PRAppl,EB_IHL_YIG_2021}. An exchange springlike characteristic is found in YIG film due to the spiral arrangement of the magnetic moments \cite{APLEva,Synthetic-antiferro_PRAppl,EB_IHL_YIG_2021}. The FMR measurement and the extracted value of $\Delta 4\pi M_{eff}$ reflect inhomogeneous distribution of $4 \pi M_{eff}$ in YIG-GdIG bilayer system. The argument of Bloch domain-wall-like spiral arrangement of moments is conceivable, as this arrangement between the adjacent layers lowers the exchange interaction energy\cite{IHL_LSMO}. To further substantiate the presence of an interfacial AFM exchange coupling leading to spin inhomogeneity at YIG-GdIG interface, we performed in-plane $\phi_{H}$ variation of $H_{r}$ on a relatively thin YIG film ($\sim 100$ nm with growth conditions leading to the formation of a GdIG interfacial layer\cite{EB_IHL_YIG_2021}). Fig. \ref{Fig:IP_Hr_Phi_H}(b) shows prominent feature of unidirectional anisotropy due to AFM exchange coupling in 100 nm thick film. It is evident that the interfacial layer exchange couples with the rest of the YIG film and leads to a unidirectional anisotropy. We observe that the interfacial exchange coupling may cause $\Delta H$ broadening and enhanced $\alpha$ due to spin inhomogeneity at the YIG-GdIG interface, even in a 200 nm thick YIG film.

\section{Conclusions}

The effects of spin inhomogeneity at the YIG and growth-induced GdIG interface on the magnetization dynamics of a 200 nm thick YIG film is studied extensively using ferromagnetic resonance technique. The Gilbert damping is almost two orders of magnitude larger ($\sim 1.2 \times 10^{-3}$) than usually reported in YIG thin films. The out-of-plane angular dependence of $\Delta H$ shows an unusual behaviour which can only be justified after considering extrinsic mechanism in combination with the Gilbert contribution. The extracted parameters from the $\Delta H$ vs $\theta_{H}$ simulation are, (i) $\alpha = 1.3 \times 10^{-3}$ from Gilbert term; (ii) $\Delta 4 \pi M_{eff} = 58$ Oe and $\Delta \theta_{H} = 0.29^{o}$ from the inhomogeneity in effective magnetization and anisotropy axes, respectively; (iii) $\Gamma_{i}^{0} = 1.3$ Oe  from TMS. The TMS strength $\Gamma$ is not so appreciable, indicating high quality thin film with insignificant defect sites. The AFM exchange coupling between YIG and the interfacial GdIG layer causes exchange springlike behaviour of the magnetic moments in YIG, leading to a large $\Delta 4 \pi M_{eff}$. The presence of large $\Delta 4 \pi M_{eff}$ impels the quick dragging of the precessional motion towards equilibrium. A unidirectional behaviour is observed in the in-plane angular variation of resonance field due to the presence of an exchange anisotropy. This further reinforces the spin inhomogeneity at the YIG-GdIG interface due to the AFM exchange coupling. \\

\textbf{ACKNOWLEDGEMENTS}

We gratefully acknowledge the research support from IIT Kanpur and SERB, Government of India (Grant No. CRG/2018/000220). RK and DS acknowledge the financial support from Max-Planck partner group. ZH acknowledges financial support from Polish National Agency for Academic Exchange under Ulam Fellowship. The authors thank Veena Singh for her help with the angular dependent FMR measurements. \\

\bibliography{Reference_SWRYIG}
\end{document}